\documentclass[10pt,a4paper]{article}
\usepackage{jheppub_kim}
\usepackage{pdflscape}
\usepackage{amsmath}
\usepackage{amssymb}
\usepackage{dcolumn}
\usepackage{bm}
\usepackage{color}
\usepackage{amsmath}
\usepackage{breqn}
\usepackage{graphicx}
\usepackage{epsfig}
\usepackage{caption,subcaption}
\usepackage{amsfonts}
\usepackage{graphicx}
\usepackage{parskip}
\usepackage{dcolumn}
\UseRawInputEncoding

\begin{document}

\title{Ricci-Non-metric Holographic Dark Energy}

\author[a]{Sukanta Bhunia,}

\author[b]{Prabir Rudra}

\author[c]{Abhijit Lahiri}

\affiliation[a,c]{Department of Mathematics, Jadavpur University, Kolkata-700 032, India.}

\affiliation[a,b]{Department of Mathematics, Asutosh College,
Kolkata-700 026, India.}

\emailAdd{bhuniasukanta@gmail.com}
\emailAdd{prudra.math@gmail.com}
\emailAdd{lahiriabhijit2000@gmail.com}

\abstract{In this work we propose a holographic dark energy model inspired by the Ricci scalar $R$ and non-metricity scalar $Q$. The infrared cutoff of the holographic dark energy is described by a combination of the Ricci scalar and the non-metricity scalar. Such a structure is very helpful because the evolution is entirely dependent on the universe's current characteristics rather than its previous or future characteristics. Additionally, the theory becomes more fundamental due to the application of invariants. We have built the model and examined its cosmological characteristics. The behavior of several cosmological parameters, including the density parameter, equation of state parameter, and deceleration parameter, is examined. It is observed that the holographic dark energy model may depict every cosmological epoch, beginning with radiation in the early universe, moving on to matter, and concluding with the dark energy-dominated period in later times. Depending on the parameter spaces selected, the equation of state parameter demonstrates that the model may have quintessence nature, phantom-divide crossing, or even phantom nature in future times.}

\maketitle

%%%%%%%%%%%%%%%%%%%%%%%%%%
\section{Introduction}
%%%%%%%%%%%%%%%%%%%%%%%%%%
For the past twenty years, the late cosmic acceleration has been a contentious topic of discussion. Recent advances in observational cosmology, including large-scale structure observations (Koivisto and Mota 2006, Daniel 2008, Nadathur et al 2020), Cosmic Microwave Background Radiation (CMBR) (Spergel et al 2003), and type Ia supernovae (Perlmutter et al 1999, Riess et al 1998, Riess et al 2004), confirm this feature of the universe. For more than a century, general relativity (GR) has been widely acknowledged as the most successful fundamental gravitational theory for explaining the large-scale structure of the universe. The matter source and the FRW spacetime in cosmology provide exact solutions for the scale factor $a(t)$, which helps us comprehend the universe's expansion. Furthermore, the inclusion of a cosmological constant into the Einstein field equation (EFE) allows for the explanation of the present rapid expansion. The universe's late-time cosmic acceleration can be explained by this very simple model, called the $\Lambda$CDM model, which fits the observational data. Unfortunately, this beautiful model is restricted by the problems with vacuum energy (Copeland et al 2006, Nojiri and Odintsov 2007) fine-tuning. As such, it is imperative to consider alternative approaches such as dark energy (DE).

The literature contains a variety of DE candidates. Notable examples include scalar field models and Chaplygin gas models \cite{cg1, cg2, cg3, cg4, cg5}. According to the holographic principle, which derives from black hole thermodynamics, a system's entropy is determined by its area rather than its volume \cite{holo1, holo2}. This holographic principle, which is also related to string theory \cite{holo1, string1}, has led to the development of holographic dark energy (HDE) \cite{hde1, hde2, hde3}. The greatest distance that may be covered by a quantum field theory is known to have connections with an ultraviolet (UV) cutoff \cite{uv1}. The vacuum energy, which is a type of DE with a holographic origin, is directly related to this UV cutoff. For a comprehensive overview of HDE, readers might consult \cite{hderev1}. Both the basic and expanded versions of HDE have been the subject of much research, and the model has proven to be highly effective over time \cite{hde4, hde5, hde6, hde7, hde8, hde9, hde10, pamm}. The adaptability of HDE models with observational data has been one of its main advantages \cite{data1, data2, data3}.

The HDE density is known to be proportional to the inverse square of the infrared cutoff $L$, which is determined by
\begin{equation}\label{densityrhde}
\rho_{DE}=\frac{3c}{\kappa^{2}L^{2}}
\end{equation}
where $\kappa^{2}$ is the gravitational constant and $c$ is a
parameter. Here we have considered the Planck mass $M_{p}$ as unity. However, there is no consensus on the appropriate infrared (IR) cutoff for the cosmic application of the holographic principle. The particle horizon and the Hubble radius, which are unable to propel the cosmic acceleration, are the most popular options \cite{hubh}. Last but not least, the future event horizon is appropriate for the situation and can function as the infrared cutoff \cite{hde3}. Despite being a good fit for the situation, this decision has certain logical issues. The future evolution of the dark energy density actually determines its present value, which is a difficult idea to comprehend. Therefore, more efforts have been made to identify modified holographic dark energy models in which the DE density is dependent on past and present evolution rather than future evolution. Agegraphic dark energy refers to models in which the IR cutoff can be determined by quantities based on the universe's historical aspects; in this case, the age of the universe or the conformal time \cite{age1, age2, age3} provides the IR cutoff. The inverse square root of the Ricci scalar $R$ is used as the IR cutoff in the model that relies on the current universe's evolution \cite{age4}. We refer to this as the Ricci holographic dark energy (RHDE). The additional benefit of using the Ricci scalar as the infrared cutoff is that the dark energy evolution is controlled by a gravitational invariant, which is crucial to gravitational theories.  Furthermore, its evolution is only dependent on the universe's current characteristics, which is a far more sound theoretical idea.  In addition, there are intriguing cosmological uses for Ricci holographic dark energy \cite{age4, ricc1, ricc2, ricc3}. In refs. \cite{od1, od2, od3, od4, od5} some generalized characteristics of holographic dark energy are presented. Moreover \cite{od6, od7} lists a few phenomenological issues with minimal holographic dark energy.

Since invariants are essential to physics, it follows that we can represent the infrared cutoff using different invariants, leading to the creation of new HDE forms. The Gauss-Bonnet (GB) invariant $G$ would be the most straightforward example of such an extension. In order to solve this issue, Saridakis in \cite{rgbhde} described the IR cutoff using a combination of the GB invariant and the Ricci scalar $R$, as shown below,
\begin{equation}
\frac{1}{L^{2}}=-\alpha R+\beta \sqrt{|G|}
\end{equation}
where $\alpha$ and $\beta$ are constants. This version of HDE was referred to as the Ricci-Gauss Bonnet HDE (RGBHDE), and it displayed fascinating cosmological characteristics, encompassing the complete thermal evolution of the universe, starting from the radiation era, progressing through the matter era, and ultimately leading to the phase dominated by dark energy. Rudra in \cite{cubic1} used the cubic invariant $P$ to propose the Ricci-cubic HDE (RCHDE), which exhibited very interesting cosmological perspectives. In this model, the IR cutoff was described by,
\begin{equation}
\frac{1}{L^{2}}=-\gamma R+\lambda P^{1/3}
\end{equation}
where $\gamma$ and $\lambda$ are constants. Further development to this HDE model has been done in \cite{cubic2}. Although different in formulation and complexity, both the RGBHDE and RCHDE, basically use the Ricci scalar $R$ and its higher orders to describe the IR cutoff. So it is understandable that they differ analytically due to their different orders of curvature and hence possess different degrees of freedom. But since both of them are constructed using the Ricci $R$, it may seem that they possess similar cosmological flavors as RHDE. Naturally, a question arises: Do we have a gravitational invariant that is different from the Ricci scalar from the topological aspect? The search takes us to the gravitational trinity. We see that apart from the Ricci scalar, we can also use the non-metricity scalar $Q$ as the gravity Lagrangian to get an alternative gravity theory.

GR and its extensions like the $f(R)$, $f(G)$, $f(P)$ gravity theories share a common trait, which is the underlying Riemannian geometry (formulated in the Riemann metrical space) situated at the core of all these classical theories. Even the torsion equivalent of GR (TEGR) and its extension to teleparallel gravity is metrical in nature, which means that these theories are metric theories of gravitation satisfying a vanishing covariant derivative of the metric tensor. Given that these theories are incompatible at some scales, one could wonder if any of the long-standing disputes in these classical theories could be settled if the underlying geometry were replaced with a far more universal geometric framework. Weyl \cite{weyl1}, whose primary goal was the mathematical unification of gravity and electromagnetism, made such a bold attempt. The Levi-Civita connection, which is compatible with the metric, is known to be the fundamental tool for contrasting vector lengths in the geometry of curved space. Weyl's theory uses a novel method that uses two connections: one transmits the length information of a vector while the other controls its direction during parallel transit. The non-vanishing covariant derivative of the fundamental tensor, which leads to non-metricity (NM) $Q$, a new geometric quantity, is the most prominent aspect of this. It is thought that the elementary geometric parameter required for various gravitational linkages is represented by the non-metricity $Q$. Symmetric teleparallel gravity (STG) is the name given to this theory \cite{q1}. Non-metric gravity and coincident GR are other names for $f(Q)$ gravity \cite{q2}. In \cite{q3, q4}, the cosmology of $f(Q)$ gravity and its empirical restrictions were examined. The STG framework has undergone several improvements in the last few decades \cite{q5, q6, q7, q8, q9}.

Herein, our primary objective is to propose and explore the dynamics of a holographic dark energy inspired by the NM scalar $Q$ and the Ricci scalar $R$. We will use the NM scalar coupled to the Ricci scalar as a measure of the size of the universe as is required for holographic dark energy and then proceed to get the energy density. Considering a Friedmann-Robertson-Walker (FRW) spacetime we plan to explore the cosmological implication of the HDE model. The work is structured as follows: The HDE model from the NM scalar $Q$ and Ricci scalar $R$ is proposed in section II. In section III we we study the cosmological evolution of the new HDE model. The paper ends with a conclusion and discussion in section IV.

%%%%%%%%%%%%%%%%%%%%%%%%%%%%%%%%%%%%%%%%%%%%%%%%%%%%%
\section{Ricci-Non-metric Holographic dark energy (RNMHDE)}
%%%%%%%%%%%%%%%%%%%%%%%%%%%%%%%%%%%%%%%%%%%%%%%%%%%%%
The NM scalar is defined as
\begin{equation}\label{nm1}
Q=-\frac{1}{4}Q_{\alpha\beta\gamma}Q^{\alpha\beta\gamma}+\frac{1}{2}Q_{\alpha\beta\gamma}Q^{\gamma\beta\alpha}+\frac{1}{4}Q_{\alpha}Q^{\alpha}-\frac{1}{2}Q_{\alpha}\Tilde{Q}^{\alpha},
\end{equation}
where
\begin{equation}\label{nm2}
Q_{\alpha}\equiv Q_{\alpha\mu}^{\mu},
\end{equation}
\begin{equation}\label{nm3}
\Tilde{Q}^{\alpha}\equiv Q_{\mu}^{\mu\alpha}
\end{equation}
and the NM tensor as,
\begin{equation}\label{nm4}
Q_{\alpha\mu\nu}\equiv\nabla_{\alpha}g_{\mu\nu}.
\end{equation}

The flat, homogeneous, and isotropic Friedmann-Lemaitre-Robertson-Walker (FLRW) metric is given as
\begin{equation}
ds^{2}=-dt^{2}+a^{2}(t)\left[dr^{2}+r^{2}\left(d\theta^{2}+\sin^{2}{\theta} d\phi^{2}\right)\right]
\end{equation}
where $a(t)$ is the cosmological scale factor with $t$ denoting the cosmic time. The NM scalar in the FLRW spacetime is given by,
\begin{equation}\label{symm}
Q=6H^{2}
\end{equation}
In this work, we will use the NM scalar to describe the IR cutoff of the HDE model. The Ricci scalar has dimensions of squared inverse length ([$L^{-2}]$). While using a combination of invariants, we have to make sure that they are of the same dimensions. The NM scalar $Q$ is constructed from the derivatives of the metric through the non-metricity tensor and its contractions. We express the cut-off as
\begin{equation}\label{ircut}
\frac{1}{L^{2}}=-\alpha R+\beta Q
\end{equation}
where the constants $\alpha$ and $\beta$ are model parameters. For $\beta=0$ we get the Ricci HDE, and for $\alpha=0$ we get pure non-metric HDE. Using eqn.(\ref{ircut}) in eqn.(\ref{densityrhde}) we get the energy density of our proposed HDE model as
\begin{equation}\label{rhohde}
\rho_{DE}=\frac{3}{\kappa^{2}}\left(-\alpha R+\beta Q\right)
\end{equation}
Here, the constant $c$ has been absorbed in the model parameters for convenience. For the flat FLRW geometry the Ricci scalar is given by
\begin{equation}\label{ricci1}
R=-6\left(2H^{2}+\dot{H}\right)    
\end{equation}

Using eqns.(\ref{symm}) and (\ref{ricci1}) we get the energy density as,
\begin{equation}
\rho_{DE}=\frac{18}{\kappa^{2}}\left[\left(2\alpha+\beta\right)H^2+\alpha \dot{H}\right]
\end{equation}
The FLRW equation is given by
\begin{equation}\label{frw11}
3H^{2}=\kappa^{2}\left(\rho_{m}+\rho_{DE}\right)
\end{equation}
where $\rho_{m}$ is the energy density of matter, whose equation of state is given by $w_m=\frac{p_{m}}{\rho_{m}}$, where $p_{m}$ is the pressure of matter. The matter sector obeys the following conservation equation
\begin{equation}\label{cons}
\dot{\rho}_{m}+3H\left(\rho_{m}+p_{m}\right)=0
\end{equation}
The full evolution history of the universe can be determined using the above equations (\ref{frw11}) and (\ref{cons}), provided $w_m$ is known. In this work we will consider pressure-less matter $p_{m}=0$, which leads to $w_{m}=0$. Using the conservation equation, for pressure-less matter we get $\rho_{m}=\rho_{m0}a^{-3}$, where $\rho_{m0}$ is the present density of matter. Using the matter energy density and the HDE energy density in the FLRW equation, we will get a differential equation in terms of the scale factor $a$. Solving this we get the complete evolution of the universe. We will study this cosmological evolution of the universe filled with RNMHDE in the following section.

%%%%%%%%%%%%%%%%%%%%%%%%%%%%%%%%%
\section{Cosmological Evolution}
%%%%%%%%%%%%%%%%%%%%%%%%%%%%%%%%%
To perform a comprehensive cosmological study we will introduce the following dimensionless density parameters
\begin{equation}\label{dimless}
\Omega_{m}\equiv\frac{\kappa^{2}\rho_{m}}{3H^{2}},  ~~~~~~~~~~~~~~~~\Omega_{DE}\equiv\frac{\kappa^{2}\rho_{DE}}{3H^{2}}
\end{equation}
Using the above parameters the FLRW equation becomes 
\begin{equation}
\Omega_{m}+\Omega_{DE}=1    
\end{equation}
For pressureless dust we get $\Omega_{m}$ as
$\Omega_{m}=\Omega_{m0}H_{0}^{2}/H^{2}a^{3}$, where $\Omega_{m0}$
is the present value of $\Omega_{m}$ given by
$\Omega_{m0}=\kappa^{2}\rho_{m0}/3H_{0}^{2}$. $H_{0}$ is the present value of the Hubble function. Using this result we can
easily get the Hubble function as below,
\begin{equation}\label{hub}
H=\frac{H_{0}\sqrt{\Omega_{m0}}}{\sqrt{a^{3}\left(1-\Omega_{DE}\right)}}
\end{equation}
Here we will use $x=\ln a$ as the independent variable.
Differentiating eqn.(\ref{hub}) we have,
\begin{equation}\label{hdot}
\dot{H}=-\frac{H^{2}}{2\left(1-\Omega_{DE}\right)}\left[3\left(1-\Omega_{DE}\right)-\Omega_{DE}'\right]
\end{equation}
where prime represents derivative with respect to $x$. Using the above equation in eqn.(\ref{ricci1}) we get
\begin{equation}
R=-3H^{2}\left(1+\frac{\Omega_{DE}'}{1-\Omega_{DE}}\right)
\end{equation}
Using the expressions for $R$ and $Q$ in eqn.(\ref{rhohde}) we get,
\begin{equation}
\rho_{HDE}=\frac{9H^{2}}{\kappa^{2}}\left(\frac{\alpha\Omega_{DE}'}{1-\Omega_{DE}}+\alpha+2\beta\right)
\end{equation}
Putting the above expression for energy density in eqn.(\ref{dimless}) we get the dimensionless density parameter as
\begin{equation}
\Omega_{DE}-3\left(\frac{\alpha\Omega_{DE}'}{1-\Omega_{DE}}+\alpha+2\beta\right)=0
\end{equation}
We obtain an exact solution of the above equation as
\begin{equation}\label{exactsol}
\Omega_{DE}=3\left(\alpha+2\beta\right)+\frac{\left(3\alpha+6\beta-1\right)\exp{\left[\frac{x}{3\alpha}+C\right]}}{\exp{\left[\frac{\left(\alpha+2\beta\right)\left(x+3C\alpha\right)}{\alpha}\right]}-\exp{\left[\frac{x}{3\alpha}+C\right]}}
\end{equation}
where $C$ is the constant of integration. Eqn.(\ref{exactsol}) gives the evolution of the RNMHDE in terms of the logarithm of the scale factor. This can be evaluated in terms of redshift using the relations $z=\frac{a_{0}}{a-1}$ and $x=\ln{a}=-\ln{\left(1+z\right)}$. Eqn.(\ref{exactsol}) can be used in eqn.(\ref{hub}) to obtain the Hubble parameter. Then the resulting expression may be integrated to obtain the scale factor.

%%%%%%%%%%%%%%%%%%%%%%%%%%%%%%%%%%%%%%%%%%%%%%%%%%%%%%%%%%%%%
\begin{figure}[hbt!]
\begin{center}
\includegraphics[height=2.5in]{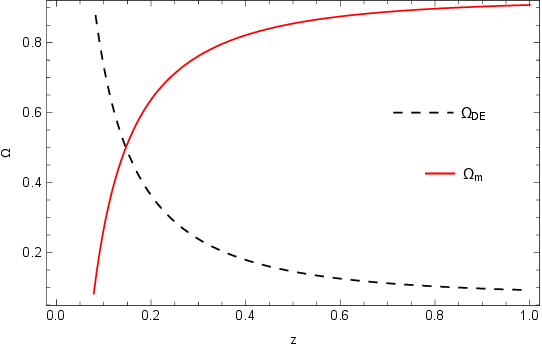}
\caption{Plot of the density parameters $\Omega_{DE}$ and $\Omega_{m}$ with respect to $z$. Other parameters are taken as $\alpha=-0.4,\beta=-0.6, H_{0}=69$. The constant of integration $C$ is fixed in order to obtain $\Omega_{DE}(z=0)\equiv\Omega_{DE0}\approx 0.68$.}
\label{fig1}
\end{center}
\end{figure}
%%%%%%%%%%%%%%%%%%%%%%%%%%%%%%%%%%%%%%%%%%%%%%%%%%%%%%%%%%%%%

The dimensionless density parameters $\Omega_{DE}(z)$ and $\Omega_{m}(z)=1-\Omega_{DE}(z)$ as functions of the redshift $z$ are shown in Fig.(\ref{fig1}). We observe that dark energy predominates over the matter sector as the universe evolves, which is consistent with established findings.

The equation of state (EoS) parameter $w_{DE}=p_{DE}/\rho_{DE}$, where $p_{DE}$ and $\rho_{DE}$ are the pressure and density of HDE, respectively, is another crucial parameter for the dark energy paradigm. Since matter is conserved in accordance with equation (\ref{cons}), non-metric HDE will be conserved according the relation
\begin{equation}
\dot{\rho}_{DE}+3H\left(1+w_{DE}\right)\rho_{DE}=0
\end{equation}
Using eqn.(\ref{rhohde}) in the above DE conservation equation we get,
\begin{equation}\label{wdee}
w_{DE}=-1+\frac{1}{3H^{2}\Omega_{DE}}\left(\alpha R'-\beta Q'\right)
\end{equation}
Now differentiating (\ref{ricci1}) and $Q=6H^2$ and using the expression (\ref{hdot}) we get
\begin{equation}\label{rdash}
R'=3H^{2}\left(3+\frac{2\Omega_{DE}'}{1-\Omega_{DE}}-\frac{2(\Omega_{DE}')^{2}}{\left(1-\Omega_{DE}\right)^{2}}-\frac{\Omega_{DE}''}{1-\Omega_{DE}}\right)    
\end{equation}
and 
\begin{equation}\label{qdash}
Q'=-6H^{2}\left(3-\frac{\Omega_{DE}'}{1-\Omega_{DE}}\right)
\end{equation}
Using eqns.(\ref{rdash}) and (\ref{qdash}) in eqn.(\ref{wdee}) we get the EoS paramneter as
\begin{equation}
w_{DE}=-1+\Omega_{DE}^{-1}\left[3\left(\alpha+2\beta\right)+2\left(\alpha-\beta\right)\frac{\Omega_{DE}'}{1-\Omega_{DE}}-\alpha\left(\frac{2(\Omega_{DE}')^{2}}{\left(1-\Omega_{DE}\right)^{2}}+\frac{\Omega_{DE}''}{1-\Omega_{DE}}\right)\right]
\end{equation}

\begin{figure}[hbt!]
\begin{center}
\includegraphics[height=2.5in]{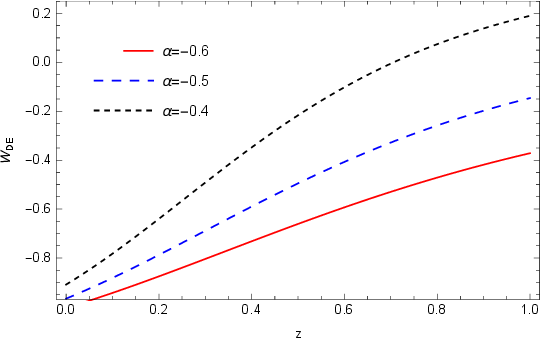}
\caption{Plot of EoS parameter $w_{DE}$ with respect to $z$ for different values of $\alpha$. We have taken $\beta=-0.6$. The constant of integration $C$ is fixed in order to obtain $\Omega_{DE}(z=0)\equiv\Omega_{DE0}\approx 0.68$.}
\label{fig2}
\end{center}
\end{figure}

\begin{figure}[hbt!]
\begin{center}
\includegraphics[height=2.5in]{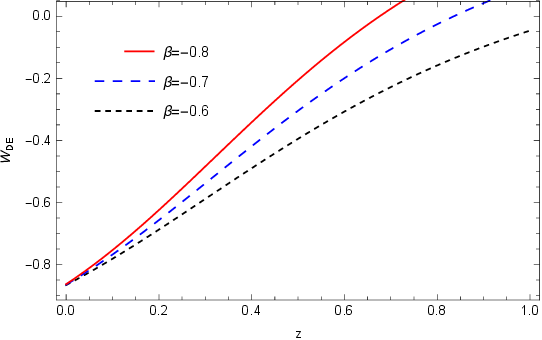}
\caption{Plot of EoS parameter $w_{DE}$ with respect to $z$ for different values of $\beta$. We have taken $\alpha=-0.5$. The constant of integration $C$ is fixed in order to obtain $\Omega_{DE}(z=0)\equiv\Omega_{DE0}\approx 0.68$.}
\label{fig3}
\end{center}
\end{figure}

The EoS parameter is shown against redshift in Figs.(\ref{fig2}) and (\ref{fig3}). In Fig.(\ref{fig2}) plots are generated for different values of $\alpha$ and in Fig.(\ref{fig3}) the plots are generated for different values of $\beta$. These plots clearly show that the universe is currently in a dark energy dominated phase ($w_{DE}<-1/3$) at $z=0$. Indeed, the trajectories descend far into the DE-dominated era, where there is a good chance of phantom-divide crossing ($w_{DE}<-1$). Moreover they show that the EoS parameter depends considerably on the model parameters $\alpha$ and $\beta$. In Fig.(\ref{fig2}) we see that as the values of $\alpha$ decrease the trajectories are lowered, showing decreased values of $w_{DE}$. This dependence on $\alpha$ is highly pronounced in the early times and decays as we approach $z\to 0$. Near the present times the trajectories tend to merge, thus alleviating the $\alpha$-dependence. Similar trend is observed in Fig.(\ref{fig3}) where the dependence on $\beta$ has been demonstrated. Due to the presence of the extreme negative pressure in the late universe, the dependence of the EoS parameter on the model parameters decreases as the trajectories are pulled towards the negative region. In these plots, we have fixed the integration constant $C$ in such a way so that we obtain $\Omega_{DE}(z=0)\equiv\Omega_{DE0}\approx 0.68$, which is the observationally favored value \cite{planckk}.

%%%%%%%%%%%%%%%%%%%%%%%%%%%%%%%%%%%%%%%%%%%%%%%%%%%%%%%%%%%%%%%
%%%%%%%%%%%%%%%%%%%%%%%%%%%%%%%%%%%%%%%%%%%%%%%%%%%%%%%%%%%%%%%

%%%%%%%%%%%%%%%%%%%%%%%%%%%%%%%%%%%%%%%%%%%%%%%%%%%%%%%%%%%%%%%
%%%%%%%%%%%%%%%%%%%%%%%%%%%%%%%%%%%%%%%%%%%%%%%%%%%%%%%%%%%%%%%

\begin{figure}[hbt!]
\begin{center}
\includegraphics[height=2.5in]{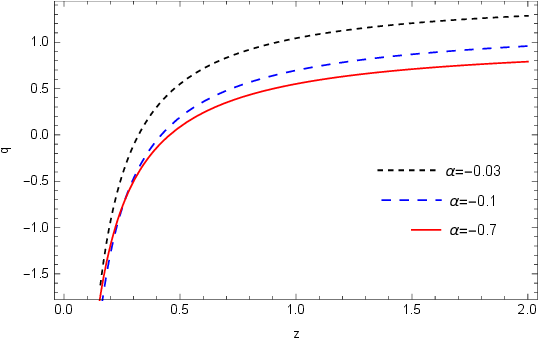}
\caption{Plot of deceleration parameter $q$ with respect to $z$ for different values of $\alpha$. We have taken $\beta=-0.6$. The constant of integration $C$ is fixed in order to obtain $q\approx 0$ around $z=0.45$ where the transition occurs.}
\label{fig4}
\end{center}
\end{figure}

\begin{figure}[hbt!]
\begin{center}
\includegraphics[height=2.5in]{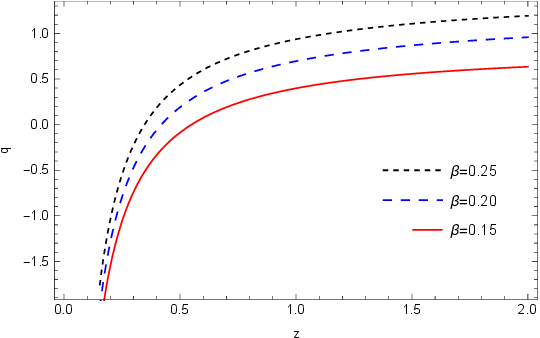}
\caption{Plot of deceleration parameter $q$ with respect to $z$ for different values of $\beta$. We have taken $\alpha=-0.5$. The constant of integration $C$ is fixed in order to obtain $q\approx 0$ around $z=0.45$ where the transition occurs.}
\label{fig5}
\end{center}
\end{figure}

Finally we can introduce the deceleration parameter for our dark energy model as
\begin{equation}
q=-1-\frac{\dot{H}}{H^2}=\frac{1}{2}+\frac{3}{2}\left(w_{m}\Omega_{m}+w_{DE}\Omega_{DE}\right)
\end{equation}
We have generated the plots for the deceleration parameter $q$ against redshift $z$ in the Figs.(\ref{fig4}) and (\ref{fig5}). In Figs.(\ref{fig4}) and (\ref{fig5}) we have generated the trajectories for varying $\alpha$ and $\beta$ respectively. In both the plots we see that there is a transition from decelerated to accelerated regime as the universe evolves towards the present time, i.e, $z\to 0$. The transition occurs around $z\approx 0.45$ which is the observationally favored redshift value. In Fig.(\ref{fig4}) we see that as $\alpha$ decreases the value of $q$ also lowers. The dependence on $\alpha$ is more pronounced at early universe and decreases as we move towards present time. Similar type of dependence is observed for $\beta$ in Fig.(\ref{fig5}). In both the figures we see that near the present time $z\approx 0$ the trajectories almost merge and become asymptotic. This shows that the universe has a tendency to dive deep down in the accelerated regime in the present and future times. In these plots, we have fixed the integration constant $C$ in such a way so that we obtain $q\approx 0$, at around $z=0.45$  which is the observationally favored value for the transition. Dynamical system analysis is necessary to comprehend the stability of the different asymptotic solutions and to obtain additional insights into the future evolution of the universe given by Ricci-non metric HDE. Such a study is outside the purview of the current project and may be undertaken in the future.

%%%%%%%%%%%%%%%%%%%%%%%%%%%%%%%%%%%%%%%%%%%%%%%%%%%%%%%%%%%%%%%
%%%%%%%%%%%%%%%%%%%%%%%%%%%%%%%%%%%%%%%%%%%%%%%%%%%%%%%%%%%%%%%

%%%%%%%%%%%%%%%%%%%%%%%%%%%%%%%%%%%%%%%%%%%%%%%%%%%%%%%%%%%%%%%
%%%%%%%%%%%%%%%%%%%%%%%%%%%%%%%%%%%%%%%%%%%%%%%%%%%%%%%%%%%%%%%

If we consider $\beta=0$, we get the usual Ricci dark energy and the corresponding equations are
\begin{equation}
\Omega_{DE|Ricci}=\frac{3\alpha e^{\left(3\alpha-1\right)\left(\frac{x}{3\alpha}+C_{1}\right)}-1}{e^{\left(3\alpha-1\right)\left(\frac{x}{3\alpha}+C_{1}\right)}-1}
\end{equation}
which is similar to the forms obtained in \cite{rgbhde, cubic1}. Here $C_{1}$ is the constant of integration. Using this the Hubble parameter from the Ricci contribution is given by
\begin{equation}
H_{Ricci}=H_{0}\sqrt{\Omega_{m0}}e^{-2x}\left[\frac{\left(3\alpha-1\right)e^{3\alpha C_{1}}}{e^{C_{1}+\frac{x}{3\alpha}}-e^{x+3\alpha C_{1}}}\right]^{-1/2}
\end{equation}
The corresponding energy density is given by
\begin{equation}
\rho_{DE|Ricci}=\frac{3H_{0}^{2}\Omega_{m0}\left(3\alpha e^{\left(3\alpha-1\right)\left(\frac{x}{3\alpha}+C_{1}\right)}-1\right)\left(e^{\frac{x}{3\alpha}-3\alpha C_{1}+C_{1}}\right)}{\kappa^{2}\left(1-3\alpha\right)e^{4x}}
\end{equation}
The corresponding EoS parameter from the Ricci contribution is given by
\begin{equation}
w_{DE|Ricci}=\frac{3\alpha-1}{9\alpha\left(1-3\alpha e^{\left(3\alpha-1\right)\left(\frac{x}{3\alpha}+C_{1}\right)}\right)}    
\end{equation}
In the similar manner we can obtain the contributions of pure non-metric HDE by putting $\alpha=0$. The dimensionless density parameter is given by
\begin{equation}
\Omega_{DE|nm}=6\beta
\end{equation}
We see that there is a constant contribution from the non-metricity sector to the density parameter. The corresponding Hubble parameter is given by
\begin{equation}
H_{nm}=H_{0}\sqrt{\Omega_{m0}}\left(1-6\beta\right)^{-1/2}e^{-3x/2}
\end{equation}
The corresponding energy density of the pure non-metric HDE can be given by
\begin{equation}
\rho_{DE|nm}=\frac{18H_{0}^{2}\Omega_{m0}\beta}{\left(1-6\beta\right)\kappa^{2}}e^{-3x}
\end{equation}
From the above expressions we see that Hubble parameter and the energy density falls exponentially as the universe expands. Finally, the EoS parameter may be given by,
\begin{equation}
w_{DE|nm}=0
\end{equation}
This is a very interesting result. We see that pure non-metric HDE actually behaves like a pressure-less dust in the cosmological regime. This shows that pure non-metric HDE do not have any dark energy contribution. It is only when it is coupled with Ricci scalar to form the Ricci-Non-metric HDE, it plays a non-trivial kinematic role to the expanding scenario, providing a negative energy component.

Lastly, we will use the Ricci-non-metric HDE to explore two distinct cosmological setups. In the first scenario, we will investigate an interaction between dark matter and the Ricci-non-metric HDE. In the second scenario, we shall explain the complete thermal history of the universe by including the radiation sector in our set-up.

%%%%%%%%%%%%%%%%%%%%%%%%%%%%%%%%%%%%%%%%%%%%%%%%%%%%%%%%%%%%%%%%%%
\subsection{Interacting scenario between RNMHDE and dark matter}
%%%%%%%%%%%%%%%%%%%%%%%%%%%%%%%%%%%%%%%%%%%%%%%%%%%%%%%%%%%%%%%%%%
Since the interaction between the dark energy and dark matter sectors cannot be ruled out by logic, a generalized scenario should permit it. Additionally, the cosmic coincidence problem (which calls into question the equality of the density parameters of the dark energy and dark matter sectors despite the fact that these components have evolved according to entirely different scales) is alleviated when interaction is included in cosmological models \cite{coin1, coin2}. When the interaction term is included, the conservation equations for matter and HDE become, respectively,
\begin{equation}\label{matcon}
\dot{\rho}_{m}+3H\left(\rho_{m}+p_{m}\right)=-Q
\end{equation}
and
\begin{equation}\label{encon}
\dot{\rho}_{DE}+3H\left(1+w_{DE}\right)\rho_{DE}=Q
\end{equation}
where $Q$ represents the interaction between dark energy and dark matter sectors. In this case, a transfer from the dark matter sector to the dark energy sector is indicated by $Q>0$, and the opposite flow is shown by $Q<0$. Therefore, a positive interaction term essentially denotes the dark energy sector's expansion, while a negative interaction term denotes the dark energy sector's decline. In the literature, there are several ways to theorize a good form of the interaction term. It is clear that the interaction should be proportional to the energy density of the universe's constituents in almost all of these forms. A well motivated form of interaction widely studied in literature is $Q=\eta H \rho_{m}$, where $\eta$ represents the rate of interaction between the sectors \cite{coin1, coin2}. In the context of interaction with the dark matter sector, we continue to investigate the cosmology of Ricci-non-metric HDE. From eqn.(\ref{matcon}) we get the energy density of matter as
\begin{equation}
\rho_{m}=\rho_{m0}a^{-\left(3+\eta\right)}
\end{equation}
The Hubble parameter gets modified to
\begin{equation}\label{hub1}
H=\frac{H_{0}\sqrt{\Omega_{m0}}}{\sqrt{a^{3+\eta}\left(1-\Omega_{DE}\right)}}
\end{equation}
The time derivative of the Hubble parameter becomes,
\begin{equation}\label{hdot1}
\dot{H}=-\frac{H^{2}}{2\left(1-\Omega_{DE}\right)}\left[\left(3+\eta\right)\left(1-\Omega_{DE}\right)-\Omega_{DE}'\right]
\end{equation}
The corresponding expression for the Ricci scalar and non-metricity scalar are given by
\begin{equation}
R=-3H^{2}\left(1-\eta+\frac{\Omega_{DE}'}{1-\Omega_{DE}}\right)
\end{equation}
and
\begin{equation}
Q=6H^{2}
\end{equation}
The expression for energy density is obtained as
\begin{equation}
\rho_{DE}=\frac{9H^{2}\left(\alpha+2\beta-\alpha\eta+\frac{\alpha\Omega_{DE}'}{1-\Omega_{DE}}\right)}{\kappa^{2}}
\end{equation}
The differential equation for $\Omega_{DE}$ can be obtained as
\begin{equation}
\Omega_{DE}-3\left(\frac{\alpha\Omega_{DE}'}{1-\Omega_{DE}}+\alpha+2\beta-\alpha\eta\right)=0
\end{equation}
Solving the above equation we get the dimensionless density parameter $\Omega_{DE}$ as
\begin{equation}
\Omega_{DE}=1+\frac{e^{x+\frac{2x\beta}{\alpha}+C_{2}+3C_{2}\alpha\eta}\left(1-6\beta+3\alpha\left(\eta-1\right)\right)}{e^{x\left(\frac{1}{3\alpha}+\eta\right)+3C_{2}\left(\alpha+2\beta\right)}-e^{x+\frac{2\beta x}{\alpha}+C_{2}+3C_{2}\alpha\eta}}    
\end{equation}
where $C_{2}$ is the constant of integration. For $\eta>0$ we have a quicker realization of the dark energy domination, but for $\eta<0$, the dark energy domination is delayed.

\subsection{Evolution in the presence of radiation}
We will incorporate the radiation component into our setup and thoroughly investigate the scenario to probe the radiation-dominated early cosmos as suggested by the observations. This will help us to accurately describe the entire thermal history of the universe. The evolution equations of the Ricci-non-metric HDE in the presence of radiation will be examined in this section. We start by considering the dimensionless density parameter of radiation
\begin{equation}
\Omega_{r}=\frac{\kappa^{2}\rho_{r}}{3H^{2}}
\end{equation}
We include the radiation component in the FLRW equation (\ref{frw11}) we get $\Omega_{m}+\Omega_{DE}+\Omega_{r}=1$. The corresponding Hubble parameter becomes
\begin{equation}
H=\frac{H_{0}\sqrt{\Omega_{m0}}}{\sqrt{a^{3}\left(1-\Omega_{DE}-\Omega_{r}\right)}}
\end{equation}
The time derivative of the Hubble parameter is given by
\begin{equation}
\dot{H}=-\frac{H^{2}}{2\left(1-\Omega_{DE}-\Omega_{r}\right)}\left[3\left(1-\Omega_{DE}-\Omega_{r}\right)-\Omega_{DE}'-\Omega_{r}'\right]
\end{equation}
The modified Ricci scalar is given by
\begin{equation}
R=-3H^{2}\left(1+\frac{\Omega_{DE}'+\Omega_{r}'}{1-\Omega_{DE}-\Omega_{r}}\right)
\end{equation}
For this case the modified differential equation for $\Omega_{DE}$ is given by
\begin{equation}
\Omega_{DE}-3\left(\frac{\alpha\left(\Omega_{DE}'+\Omega_{r}'\right)}{1-\Omega_{DE}-\Omega_{r}}+\alpha+2\beta\right)=0
\end{equation}
The modified EoS parameter is given by
\begin{equation}
w_{DE}=-1+\Omega_{DE}^{-1}\left[3\left(\alpha+2\beta\right)+2\left(\alpha-\beta\right)\frac{\Omega_{DE}'+\Omega_{r}'}{1-\Omega_{DE}-\Omega_{r}}-\alpha\left(\frac{2\left(\Omega_{DE}'+\Omega_{r}'\right)^{2}}{\left(1-\Omega_{DE}-\Omega_{r}\right)^{2}}+\frac{\Omega_{DE}''+\Omega_{r}''}{1-\Omega_{DE}-\Omega_{r}}\right)\right]
\end{equation}

%%%%%%%%%%%%%%%%%%%%%%%%%%%%%%%%%%%%%%%
\section{Discussion \& Conclusion}
%%%%%%%%%%%%%%%%%%%%%%%%%%%%%%%%%%%%%%
Inspired by the cubic curvature invariant, we have postulated the Ricci-non-metric holographic dark energy in this study. Here, the Ricci scalar and the non-metricity scalar work together to determine the infrared cutoff. This formulation has an advantage over the conventional ones (such as the Ricci holographic dark energy or the agegraphic dark energy model) in that the infrared cutoff and, hence, the holographic dark energy density are independent of the universe's past or future evolution. Another benefit is that invariants, which are crucial to the theory of gravity, provide the infrared cutoff in this case. Using a generalized method, we incorporated contributions from the non-metric invariant and the Ricci scalar of the same order in the infrared cutoff and, as a result, in the energy density of the holographic dark energy.

Initially, we formulated the model and derived the equations of the Ricci-non-metric holographic dark energy. We then studied the cosmological evolution of the universe filled with Ricci-non-metric holographic dark energy while taking into account the Friedmann metric. For both matter and dark energy, appropriate dimensionless density parameters were considered. The logarithm of the scale factor, which is directly connected to the redshift, was used to create a differential equation involving the dimensionless density parameter of the holographic dark energy. This equation admitted an exact solution, which was found. In terms of the dimensionless density parameters, we also calculated the deceleration and equation of state parameters.

The cosmological characteristics of Ricci-non-metric HDE model are quite intriguing and offer some enhanced dynamics. This is because the two curvature invariants are controlled by two model parameters. Plots for the cosmological parameters were produced, and it was observed that they complied with the universe's cosmic evolution. Additionally, it was observed that the onset of the accelerated expansion may be fine-tuned at $z\approx0.45$, which is consistent with the observational data. It was shown that during the cosmological evolution, we have a complete quintessence-like behavior in the late universe. However, looking at the trajectories, it is understandable that $\Lambda$CDM and phantom-like behavior is also possible in future times for appropriate parameter spaces. Under certain circumstances, this model also provides strong support for an asymptotic de-Sitter-like development. The standard Ricci HDE is obtained when the non-metric invariant contribution is set to zero. Furthermore, we obtain the "pure" non-metric HDE when the Ricci scalar contribution is set to zero. When we finally added radiation to our setup, it was shown that radiation, matter and dark energy epoch, can occur in sequence as desired. The late cosmic acceleration was ultimately caused when the universe entered an era dominated by dark energy. It should be noted that in order to comply with the known results, a suitable set of parameter values is used to construct the plots. By limiting the parameters with observational data, this can be well-adjusted in the future.

In the end, we state that further research is necessary to fully comprehend the nature of Ricci-non-metric HDE and, as a result, accept it as a viable dark energy candidate. We should use dynamical system analysis to investigate the global behavior of the model at late times. We should employ observational data from Type Ia supernovae (Sn Ia), baryon acoustic oscillations (BAO), the cosmic microwave background (CMB) shift parameter, and Hubble parameter observations to constrain the model parameters $\alpha$ and $\beta$. In addition, we must conduct a perturbation analysis to verify the model's stability. Future research on the Ricci-non-metric HDE model is encouraged by these essential investigations.

%%%%%%%%%%%%%%%%%%%%%%%%%%
\section*{Acknowledgments}
%%%%%%%%%%%%%%%%%%%%%%%%%%
P.R. acknowledges the Inter-University Centre for Astronomy and Astrophysics (IUCAA), Pune, India for granting visiting associateship.

%%%%%%%%%%%%%%%%%%%%%%%%%%%%%%%%%%%%%%%
\section*{Data Availability Statement}
%%%%%%%%%%%%%%%%%%%%%%%%%%%%%%%%%%%%%%%
No data was generated or analyzed in this study.

%%%%%%%%%%%%%%%%%%%%%%%%%%%%%%%%%%%%
\section*{Conflict of Interest}
%%%%%%%%%%%%%%%%%%%%%%%%%%%%%%%%%%%%
There are no conflicts of interest.

%%%%%%%%%%%%%%%%%%%%%%%%%%%%%%%
\section*{Funding Statement}
%%%%%%%%%%%%%%%%%%%%%%%%%%%%%%%
There is no funding to report for this article.

\end{document}